\documentclass[amsmath,twocolumn,superscriptaddress,prl,aps]{revtex4-1} 
\usepackage{amsthm,amsfonts,graphicx,verbatim, color}
\usepackage{bm}
\usepackage[utf8]{inputenc}
\let\oldmarginpar\marginpar
\renewcommand\marginpar[1]{\-\oldmarginpar[\raggedleft\footnotesize #1]%
{\raggedright\footnotesize #1}}

\newcommand{\be}{\begin{equation}}
\newcommand{\ee}{\end{equation}}
\newcommand{\bea}{\begin{eqnarray}}
\newcommand{\eea}{\end{eqnarray}}

\renewcommand{\epsilon}{\varepsilon}
\renewcommand{\vec}[1]{{\bf #1}}

\def\beq{\begin{equation}}
\def\eeq{\end{equation}}
\def\bea{\begin{eqnarray}}
\def\eea{\end{eqnarray}}

\begin{document}

\title{A many body localization proximity effect}
\author{Rahul Nandkishore}
\affiliation{Princeton Center for Theoretical Science, Princeton University, Princeton, New Jersey 08544, USA}
\begin{abstract}
We examine what happens when a strongly many body localized system is coupled to a weak heat bath, with both system and bath containing similar numbers of degrees of freedom. Previous investigations of localized systems coupled to baths operated in regimes where the back action of the system on the bath is negligible, and concluded that the bath generically thermalizes the system. In this work we show that when the system is strongly localized and the bath is only weakly ergodic, the system can instead  localize the bath. We demonstrate this both in the limit of weak coupling between system and bath, and in the limit of strong coupling, and for two different types of `weak' bath - baths which are close to an atomic limit, and baths which are close to a non-interacting limit. The existence of this `many body localization proximity effect' indicates that many body localization is more robust than previously appreciated, and can not only survive coupling to a (weak) heat bath, but can even destroy the bath. 
\end{abstract}
\maketitle
 
Quantum localized systems violate many of the foundational assumptions of quantum statistical physics (such as the ergodic hypothesis), and present an exciting new frontier for research \cite{Anderson}. While localization was long believed to occur mainly in systems of non-interacting particles, the recent discovery of many body localization (MBL) \cite{Mirlin, BAA, pal, imbrie} has ignited a blaze of interest in this field.  It has been realized that quantum localized systems can display a cornucopia of exotic properties, including an emergent integrability \cite{lbits, serbyn, rms-IOM}, exotic quantum states of matter \cite{LPQO, Pekker, Vosk, Kjall, Bauer, Bahri, lspt, qhmbl, vasseur, potter}, and unexpected behavior in linear \cite{gopalakrishnan} and non-linear \cite{khemani} response. These properties not only dramatically revise our understanding of quantum statistical physics, but also offer a new route to dissipationless quantum technologies.  A summary of progress in this field can be found in the review article \cite{arcmp}. 

Most works on MBL have focused on perfectly isolated quantum systems. Experimental systems, however, are always coupled (however weakly) to a thermalizing environment. The behavior of many body localized systems coupled to a thermalizing environment was first examined in \cite{ngh, jnb, gn}, in the limit where back action on the bath was negligible and the bath could be treated as Markovian. In this limit, it was argued that an arbitrarily weak coupling to a thermodynamically large bath should restore ergodicity, thermalizing the system. These results suggested that perfect MBL would be unobservable in experiments, which would see instead only signatures of proximity to a localized phase. However, these works left open the question of whether different physics could result if the localization in the system were strong, and the bath were weak. Could many body localization then survive even after coupling to a heat bath?

In this Letter, we show that when the system of interest is strongly localized, and the `heat bath' is only weakly ergodic, then MBL in the system can not only survive coupling to the bath, but can even localize the heat bath. We call this phenomenon a `many body localization proximity effect,' and it establishes that MBL is much more robust to coupling to an environment than was previously appreciated. It also suggests a possible explanation for the numerical results recently presented in \cite{Mukerjee, DasSarma}, which counter-intuitively observed many body localization in an interacting model, when the non-interacting limit contained a single particle mobility edge. 

The system we consider consists of a $D$ dimensional lattice which hosts two species of spinless fermions - $c$ and $d$. The $c$ fermions are present with density $n_c$ and have Hamiltonian 
\begin{equation}
H_c = \sum_{\langle ij \rangle} t_c c^{\dag}_{i} c_{j} + U c^{\dag}_i c_i c^{\dag}_j c_j + \sum_i \epsilon_i c^{\dag}_i c_i \label{Hc}
\end{equation}
where $t_c$ is the hopping, $U$ is a nearest neighbor interaction, and $\epsilon$ is a random potential, drawn from a distribution of width $W$. The width of the distribution is sufficiently large that the $c$ particles in isolation are in an MBL phase, with a localization length $\xi_c.$ We do assume that $W$ is the largest scale in this Hamiltonian, and sets the characteristic energy scale of the MBL system.  
Meanwhile, the $d$ particles are present with density $n_d$, and have Hamiltonian 
\begin{equation}
H_d = \sum_{\langle ij \rangle} t_d d^{\dag}_{i} d_{j} + \lambda d^{\dag}_i d_i d^{\dag}_j d_j \label{Hd}
\end{equation}
This Hamiltonian in isolation describes a system in an ergodic phase (the ground state will be a Fermi liquid). The coupling between $c$ and $d$ systems is taken to have the form
\begin{equation}
H_{int} = \sum_i g_i c^{\dag}_i c_i d^{\dag}_i d_i \label{Hcouple}
\end{equation}
For simplicity in this work we assume $n_c = n_d$, although varying the ratio of the two densities would also be an interesting parameter to tune in future work. We consider two models. First, we consider a model (model I) where the coefficients $g_i$ are taken from some distribution (e.g. box distribution) of width $G$. Next, we consider a model (model II) where the $g_i$ are {\it uniform}, with strength $G$. Many of the results are easier to establish with a random coupling (model I), and we thus discuss this model first, but we will subsequently show that a many body localization proximity effect can {\it also} arise with uniform couplings (model II). 

{\bf {\it Weak coupling (small $G$):}} We first discuss the behavior with a weak random coupling $g_i$ taken from a distribution of width $G \ll W$. In this regime the disorder in the $c$ sector is the primary source of randomness. A many body localization proximity effect can be shown to arise in two limits: the atomic limit $t_d \rightarrow 0$, and the non-interacting $d$ limit $\lambda \rightarrow 0$. We establish by means of perturbative expansions that this `many body localization proximity effect' survives for small (but non-zero) $t_d$ and $\lambda$ respectively, such that there exists a finite sized region of parameter space where coupling a localized system to a bath results in localization of the bath. We subsequently show that this `many body localization proximity effect' also arises if the coupling is uniform rather than random. 

{\it Bath close to an atomic limit: $t_d < G < W$.} We start by taking the limit $t_d \rightarrow 0$. In this atomic limit for the $d$ fermions, the eigenstates of the combined $c$ and $d$ system take the form
\begin{equation}
|\Psi\rangle = |\phi(\{i\})\rangle \otimes \prod_{\{i\}} d^{\dag}_{i} |0\rangle \label{eig1}
\end{equation}
i.e. $d$ fermions are present on a certain set of sites $\{i\}$, and the $c$ fermions are in a localized state, the precise wave function of which depends on the set of occupied sites $\{i\}$ in the $d$ sector. The wave function in the $c$ sector depends on $\{ i \}$ because the distribution of $d$ fermions affects the random potential seen by the $c$ fermions. We note that a random distribution of atomic-limit $d$ fermions will {\it increase} the effective random potential disorder in the $c$ sector, from $W$ to $\sqrt{W^2 + G^2}$, and thus if the $c$ fermions are localized at $G=0$ (uncoupled $c$ and $d$ sectors), they should also be localized at non-zero $G$. 

We now turn on a non-zero hopping in the $d$ fermion sector $t_d \neq 0$, and ask whether the resulting system is still localized. The hopping of a single $d$ fermion changes the interaction energy between $c$ and $d$ fermion sectors by an amount of order $G$ (we are working here with random couplings $g_i$). It may also change the interaction energy in the $d$ fermion sector by an amount of order $\lambda$. Thus, naively a single hop by a $d$ fermion takes the system `off shell' by an random number drawn from a distribution of width at least $G$. However, the hopping of a single $d$ fermion changes the potential seen by the $c$ fermions, and thus in the $c$ fermion sector acts like a `quantum quench.' We now discuss to what extent the $c$ fermion system may be able to relax to accommodate this change in energy.

The $c$ sector is in a localized phase, thus it will not be able to bring the system back precisely on shell. Rather, the $c$ fermion sector should act like a `finite sized bath', with the localization length setting the size, and $s(T)$ being the entropy density. The relevant level spacing will thus be $\sim W\exp(-s(T) \xi_c^D)$.  The hopping of a $d$ fermion, combined with an appropriate relaxation in the $c$ sector on length scales short compared to $\xi_c$, must thus take the system off shell by at least $\Delta E \approx W \exp(-s(T) \xi^D)$. 

To evaluate the convergence (or not) of the locator expansion \cite{Anderson}, the change in energy must be compared to the associated matrix element. The matrix element between the two states is 
\begin{equation}
\left(  \langle 0| d_i  \otimes \langle \phi(i)| \right) t_d d^{\dag}_i d_j \left( |\phi(j) \rangle \otimes d^{\dag}_j |0\rangle  \right) \approx t_d \langle \phi(i)|\phi(j)\rangle
\end{equation}
Where $|\phi_i\rangle$ and $|\phi(j)\rangle$ are the initial and final states in the $c$ sector. We assume that the two states differ only on length scales less than $\xi_c$ (ignoring rare long range resonances). We make the standard assumption that the eigenstates are `ergodic' on length scales less than $\xi_c$ i.e. that the initial state has similar overlap with all $\exp(s(T) \xi_c^D)$ possible final states that differ from the initial state only on length scales shorter than $\xi_c$, and also that the $\exp(s(T) \xi_c^D)$ possible overlap matrix elements have random phases. Demanding normalization of the final wave function, we then conclude that a typical overlap matrix element $\langle \phi(i)|\phi(j)\rangle \approx \exp(-\frac12 s(T)\xi^D)$

The locator expansion will converge if the typical hopping matrix element is smaller than the amount by which the hop takes the system off shell. From the estimates above, we conclude that a locator expansion in small $t_d$ will converge iff
\begin{equation}
 t_d < \min(G, W \exp(-\frac12 s(T) \xi_c^D) )
\end{equation}
As long as this condition is satisfied, the combined $c$ and $d$ systems will both be localized, with eigenstates that are `close' to the form (\ref{eig1}). Thus in this there regime is a clear MBL proximity effect where coupling a localized system to a bath ends up localizing the bath instead. When this condition is violated, the locator expansion breaks down, which {\it may} indicate delocalization. 

The above analysis may be readily generalized to uniform couplings of strength $G$ (model II). As long as the density pattern in the $c$ sector is spatially inhomogenous (which should be the case in the localized regime), the {\it effective} coupling between $c$ and $d$ sectors will be random, drawn from a distribution of width $G \delta n_c$, where $\delta n_c$ is the width of the density distribution in the $c$ sector. An analogous argument may then be constructed for a many body localization proximity effect at weak $t_d$. We caution however that in the limit of site localization of $c$ fermions $t_c/W \rightarrow 0$, with uniform couplings, the effective disorder introduced in the $d$ fermion sector is binary, and has percolating equipotential surfaces which break the locator expansion. This problem is absent with model I couplings. 


{\it A weakly interacting low dimensional bath: $\lambda \ll G \ll W, t_d$ and $D=1,2$}

In this regime traditional locator expansions \cite{Anderson} fail, but we can still use weak localization results. Let us begin by taking $\lambda \rightarrow 0$. In this limit, the $d$ fermions are described by a quadratic Hamiltonian, which takes the form
\begin{equation}
H_d =  \sum_{\langle ij \rangle} t_d d^{\dag}_{i} d_{j} + \sum_i V_i d^{\dag}_i d_i
\end{equation}
where $V_i = g_i c^{\dag}_i c_i$ is a static random variable (since the $c$ fermions are in a localized phase), and can be interpreted as a disorder potential, with the precise realization of the disorder depending on the state in which the $c$ fermions are prepared. Thus, the d-fermion Hamiltonian describes fermions hopping in a random potential. It does not particularly matter whether we work with model I couplings (random coupling constants) or model II (uniform coupling constants), since as long as the density pattern in the $c$ sector is inhomogenous the $d$ fermions will see an effective random potential anyway. We choose to work with model I for convenience and specificity. As is well known, free fermions hopping in a random potential will inevitably localize in $D=1,2$ \cite{Anderson, gangoffour}, with a localization length $\xi_d$ that is power law large in $t_d/G$ for $D=1$ and exponentially large in $t_d/G$ for $D=2$. 

The argument above is cleanest in the site localization limit $\xi_c \rightarrow 0$, when the $c-d$ coupling is diagonal in the $c$ eigenbasis and there is no appreciable back action on the $c$ system. At non-zero $\xi_c$, there will be an effective four $d$ fermion interaction mediated by the $c$ fermions. This four fermion interaction may be estimated in the manner of \cite{qhmbl} and will be of order $(G^2/W) \xi_d^{-3D/2} \exp(-2/\xi_c) $ for $\xi_c<1$ and of order $(G^2/W) \xi_d^{-3D/2}\xi_c^{2D}$ for $\xi_c >1$ (see supplement for details). We are working here in units where the lattice scale has been set to one. Since an interaction is already induced by the $c$ sector, we can also turn on the intrinsic interaction $\lambda \neq 0$. The matrix elements of this interaction in the basis of localized wave functions $|\varphi\rangle$ will be $\lambda_{\alpha \beta \gamma \delta} = \lambda \sum_{\langle ij \rangle} \varphi_{\alpha}^*(i) \varphi^*_{\beta}(j) \varphi_{\gamma}(j) \varphi_{\delta}(i) \approx \lambda \xi_d^{-D}$, where we have made use of normalization of the wavefunctions. 

The locator expansion should converge (such that the dressed eigenstates of the interacting problem are close to the eigenstates of the non-interacting problem), if the matrix elements of the interaction (both intrinsic and induced) are less than the accessible level spacing $t_d \xi_d^{-4D}$ i.e.
\begin{eqnarray}
&&\max\left(\frac{G^2}{W t_d \xi_d^{D/2}} \exp(-2/\xi_c), \frac{\lambda}{t_d}\right) \xi_d^{3D} < 1; \quad \xi_c < 1 \nonumber\\ \label{cond2}
&&\max\left(\frac{G^2 \xi_c^{2D}}{W t_d \xi_d^{D/2}} , \frac{\lambda}{t_d}\right) \xi_d^{3D} < 1; \qquad \qquad \qquad \xi_c > 1 
\end{eqnarray}
We note in particular that in one (two) dimensions, $\xi_d$ is power-law (exponentially) large in $t_d/G$. For there to be a well controlled weak localization regime, we therefore require not only that $\lambda$ is small, but also that either $\xi_c \ll 1$ or that $W \gg t_d$. We note also that the convergence criterion (\ref{cond2}) is simply an estimate obtained by considering the leading order terms in perturbation theory. A calculation to all orders along the lines of \cite{BAA} is beyond the scope of the present work.  

As long as the interactions (both intrinsic and induced) are sufficiently weak, the wavefunction will take the form
\begin{equation}
|\Psi\rangle = |\phi(\{\alpha\})\rangle \otimes \prod_{\{\alpha\}} d^{\dag}_{\alpha}(\phi) |0\rangle
\end{equation}
i.e. the $d$ fermions are in a product state where a set $\{\alpha\}$ of the non-interacting wave functions are occupied, the precise shape of the non-interacting wave functions depends on the state $|\phi \rangle$ in which the $c$ fermions are prepared, and the $c$-fermions are in a state $|\phi \rangle$ which is many body localized, but depends in its detailed structure on the state of the $d$-fermions.


{\it Strong coupling (large $G$)}
We now point out that a many body localization proximity effect also arises for {\it arbitrary} $\lambda$ and $t$ in the limit $G \rightarrow \infty$. We demonstrate this for model II (uniform) couplings, but it is obviously true also for model I. 

In the large $G$ limit, we have a description in terms of three species: `bound states' whereby a $c$ and $d$ fermion sit on the same site, unbound $c$s, and unbound $d$s. A bound state cannot break apart or form, because this would change the energy by an amount of order $G$, which is the largest energy scale in the problem. 

The unbound $c$ fermions are governed by a Hamiltonian of the form (\ref{Hc}), but now with certain lattice sites forbidden (the sites occupied by the $c-d$ pairs and the unpaired $d$s). Since the $c$ fermions localized with all lattice sites allowed, and the forbidding of certain lattice sites only presents an obstruction to transport, the unpaired $c$ fermions should still be localized. Meanwhile, the bound $c-d$ pairs are extremely heavy, with an effective hopping matrix element of order $t_c t_d/G \ll t_c$, and they see the same disorder potential as the $c$ fermions. Thus, if the $c$ fermions localized in a disorder potential of magnitude $W$, then the $c-d$ composites should localize also. Finally, that leaves the unpaired $d$ fermions. These live on a random lattice, obtained from the original lattice by deleting all sites on which unpaired $c$s and $c-d$ pairs are present. (This is because hopping onto an `occupied' site changes the energy by an amount of order $G$, and in the large $G$ limit this effectively makes occupied sites inaccessible). In one dimension, any finite density of deleted sites causes the `lattice' on which the unpaired $d$ fermions move to break up into finite sized segments, on which the $d$ fermions are localized. Thus, in one dimension, in the $G \rightarrow \infty$ limit, we clearly have a system of three distinct simultaneously localized species, regardless of the values of $t_d$ and $\lambda$.  

In higher dimensions too, when a sufficiently large number of sites are deleted (large $G$ limit), the lattice on which the unpaired $d$ fermions live will break into disconnected clusters, and at this point the $d$ fermions will most certainly be localized. Thus, there must be a regime in which the coupling of system and bath leads to the formation of localized species: localized $c$ fermions, localized $c-d$ composites, and $d$ fermions which live on isolated islands that are separated by regions containing either $c$ fermions or $c-d$ composites. In fact, work on quantum percolation \cite{Fehske} suggests that for non-interacting $d$ fermions (or, presumably, weakly interacting d-fermions), the localization transition {\it precedes} the percolation transition, and $d$ fermions will quantum localize even before the lattice on which they live breaks up into disconnected clusters. The density of deleted sites may be tuned by varying either the density of particles or the energy of the state (and hence the number of $c-d$ pairs), and the quantum percolation (and hence localization) transition may also be tuned in this way. 

We note that in principle there is an `exchange' process by which an unbound $c$ and a $c-d$ composite on adjacent sites could exchange position by moving the $d$ fermion, without taking the system off shell. This is really only possible if the $c$ fermions are in the site-localization limit $W \gg t_c$ (since the $c-d$ composites are always site localized in the large $G$ limit). However, in this limit of site localization for $c$ fermions, localization requires the additional condition that the set of possible energy conserving $d$ exchanges should not percolate. 

If $G$ is large but not infinite, then $d$ fermions will in principle be able to `tunnel through' the barriers presented by $c$ fermions, but virtual processes involving $c-d$ composites will then also give rise to a random onsite potential for the lattice on which the $d$ fermions live (inherited from the inhomogenous density pattern for the $c$ fermions). In low dimensions and in the non-interacting limit $\lambda \rightarrow 0$, and for sufficiently large $G$, this will be sufficient to localize the $d$ fermions, and arguments along the lines of \cite{BAA} suggest the $d$ fermions will continue to be localized for small but non-zero $\lambda$. Of course, the existence of a `many body localization proximity effect in the  limit of weakly interacting $d$ fermions is not particularly surprising, since something similar occurs for weak $G$. The truly novel feature is the existence of an MBL proximity effect at arbitrary $t_d, \lambda$ in the limit $G \rightarrow \infty$. 

 {\it Conclusion:}
 Thus, we have demonstrated that when a many body localized system is coupled to a bath, the end result can be the survival of MBL in the system, and the localization of the bath. We have demonstrated that this happens in three distinct limits: a bath close to the atomic limit that is weakly coupled to a strongly localized system, a weakly interacting bath weakly coupled to a strongly localized system, and an bath (with arbitrary bath parameters) that is {\it strongly} coupled to a localized system. The problem of localized systems coupled to baths is thus much richer when the back action on the bath is taken into account, and indeed the coupling of a localized system and a delocalized system can result in the localization of both. This may explain recent numerical results \cite{DasSarma, Mukerjee}. The development of a general theory of localized systems coupled to baths is an interesting challenge for future work. 
 
 {\bf Acknowledgements:} I thank David Huse for several illuminating discussions regarding this work. I also thank Sarang Gopalakrishnan for a useful conversation, and for feedback on an early draft. I am supported by a PCTS fellowship.

\begin{widetext}
\section{Supplement: Effective four $d$ fermion interaction mediated by $c$ fermions}
\begin{figure}
\includegraphics[width = 0.5 \columnwidth]{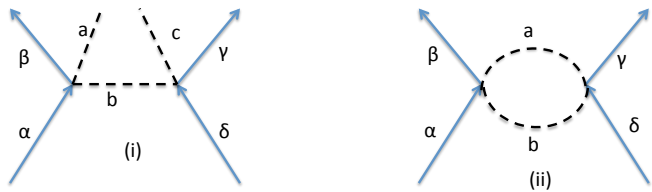}
\caption{\label{figure} Processes contributing to the mediation of an effective four $d$ fermion interaction through the $c$ fermions, at leading order in perturbation theory in weak $G$. Solid lines denote $d$ fermions and dashed lines denote $c$ fermions. The state of the $d$ fermions is labelled by Greek characters, while the state of the $c$ fermions is labelled by English characters. We assume the localization centers of the $d$ fermion wavefunctions $\phi_{\alpha, \beta, \gamma, \delta}$ are all within $\xi_d$ of each other.  }
\end{figure}
When $\xi_c \neq 0$, the coupling of $c$ and $d$ fermion sectors leads to an effective four $d$ fermion interaction. In this supplement we estimate this interaction, first for $\xi_c < 1$ and then for $\xi_c > 1$. The calculation follows \cite{qhmbl}. The two processes contributing to the effective interaction at lowest order in perturbation theory in weak $G$ are shown in Fig.\ref{figure}. The process (i) dominates and will be discussed here, but the contribution of process (ii) can be calculated similarly. 

The process Fig.\ref{figure}(i) represents an effective $T$ matrix element of the form 
\begin{equation}
T_{\alpha \beta \gamma \delta}^{a b c} = G^2 \int d^D r_1 d^D r_2 \frac{\phi^*_{\alpha}(\vec{r_1}) \phi_{\beta}(\vec{r_1}) \phi_{\gamma}(\vec{r_2}) \phi^*_{\delta}(\vec{r_2}) \psi^*_a(\vec{r_1}) \psi_b(\vec{r_1}) \psi^*_b (\vec{r_2}) \psi_c(\vec{r_2})}{\Delta E}
\end{equation}
where $\phi$ represents a $c$ fermion wavefunction, $\psi$ represents a $d$ fermion wavefunction, and $\Delta E$ represents the amount by which the intermediate state is off shell. 

If $\xi_c < 1$ then the above matrix element falls off exponentially in $|\vec{r_1} - \vec{r_2}|$ and the integral is effectively restricted to neighboring sites. Meanwhile, the scattering in the $c$ fermion sector takes us off shell by an amount of order $W$. Finally, since each site has in effect a single associated state $\psi_{a,b,c}$, the matrix element is of order $T^{abc}_{\alpha \beta \gamma \delta} \approx (G^2/W) \xi_d^{-2D} \exp(-2/\xi_c) \exp(i \phi_{abc})$, where we have assumed the $d$ wavefunctions $\phi$ are localized to a volume of radius $\xi_d$, and where $\phi_{abc}$ is a random phase. The effective induced interaction may be obtained by summing over all intermediate $c$ fermion states i.e. 
\begin{equation}
\lambda^{eff}_{\alpha \beta \gamma \delta} = \sum_{a b c} T^{abc}_{\alpha \beta \gamma \delta} \label{sum}
\end{equation}
There are $\xi_d^D$ terms contributing to the above sum (since the four $d$ fermions involved can interact through $c$ fermions anywhere in the volume where they all overlap). Each term has similar magnitude but random phase. Summing over all the terms in rms fashion gives an additional factor of $\xi_d^{D/2}$, and as a result we conclude that the effective four fermion interaction mediated by the $c$ fermions is of order $(G^2/W) \xi_d^{-3D/2} \exp(-2/\xi_c)$, as quoted in the main text.

The calculation is similar when $\xi_c > 1$. The main differences are (1) the integrals $\int d^D r_1 d^D r_2$ now go over volumes of linear dimension $\xi_c$ (2) normalization requires that $|\psi| \sim \xi_c^{-D/2}$ within $\xi_c$ of the corresponding localization center (3) the minimum energy change is $\Delta E \approx W \xi_c^{-d}$ and (4) instead of having $\xi_d^D$ terms contributing to the sum (\ref{sum}), there are $\xi_d^D \xi_c^{2D}$ contributing terms (i.e. the volume can be divided into $(\xi_d/\xi_c)^D$ cells, in each of which there are $\xi_c^{3D}$ possible choices of $a,b,c$). As a result, the effective four fermion interaction becomes of order $ (G^2/W) \xi_d^{-3D/2} \xi_c^{2D}$, again as quoted in the main text. 
\end{widetext}
\end{document}